\documentclass[sigplan,screen]{acmart}
\usepackage[label font=normal]{subfig}

\begin{document}

\title{Analyzing Performance Characteristics of PostgreSQL and MariaDB on NVMeVirt}

\author{Juhee Han}
\affiliation{%
  \institution{Seoul National University}
  \city{Seoul}
  \country{Korea}}
\email{juheehan@dbs.snu.ac.kr}

\author{Yoojin Choi}
\affiliation{%
  \institution{Seoul National University}
  \city{Seoul}
  \country{Korea}}
\email{cyj@dbs.snu.ac.kr}

\begin{abstract}
The NVMeVirt paper analyzes the implication of storage performance on database engine performance to promote the tunable performance of NVMeVirt. They perform analysis on two very popular database engines, MariaDB and PostgreSQL. The result shows that MariaDB is more efficient when the storage is slow, but PostgreSQL outperforms MariaDB as I/O bandwidth increases. Although this verifies that NVMeVirt can support advanced storage bandwidth configurations, the paper does not provide a clear explanation of why two database engines react very differently to the storage performance.

To understand why the above two database engines have different performance characteristics, we conduct a study of the database engine's internals. We focus on three major differences in Multi-version concurrency control (MVCC) implementations: version storage, garbage collection, and index management. 
We also evaluated each scheme's I/O overhead using OLTP workload. Our analysis identifies the reason why MariaDB outperforms PostgreSQL when the bandwidth is low.
\end{abstract}

\settopmatter{printacmref=false}
\setcopyright{none}
\renewcommand\footnotetextcopyrightpermission[1]{}
\pagestyle{plain}

\maketitle

\section{Introduction}
\label{sec:intro}
The NVMeVirt is a versatile software-defined virtual NVMe device. Because the NVMeVirt supports advanced storage configurations, it can be used for database engine analysis and allows us to estimate the performance of database engines on future storage devices. In NVMeVirt paper \cite{NVMeVirt:FAST23}, the authors conducted an evaluation on PostgreSQL \cite{psql} and MariaDB \cite{mariadb} with OLTP workload using sysbench \cite{sysbench}. They measured various performance metrics while running the benchmark. The result indicates that MariaDB and PostgreSQL react differently to the storage performance.

MariaDB fully utilizes the I/O bandwidth up to 500 MiB/s. However, I/O bandwidth utilization remains around 600 MiB/s even when the storage device provides higher bandwidth. On the other hand, PostgreSQL fully utilizes the I/O bandwidth up to 1,000 MiB/s, and the performance is saturated at approximately 1,800 MiB/s. The I/O bandwidth affects both database engines' performance, but PostgreSQL is much more sensitive than MariaDB. From the evaluation, the authors conclude that PostgreSQL is more promising on modern storage devices, whereas MariaDB is more efficient when the storage is low. The result verifies the tunable performance of NVMeVirt, but the problem is that the paper does not provide a clear explanation of what features of database engine internals make such differences.

In this paper, we analyze the implication of storage performance on database engine performance focusing on {\itshape database engine internals}. We aim to provide a clear explanation of why PostgreSQL is more sensitive to I/O bandwidth than MariaDB.

The contributions of this work are as follows:
\begin{itemize}
    \item We perform experiments for both OLTP and OLAP workloads and analyze the different performance characteristics (Section~\ref{sec:eval}).
    \item We analyze what differences in database engine internals make PostgreSQL more sensitive to I/O bandwidth compared to MariaDB (Section~\ref{sec:internal}).
    \item We evaluated different MVCC schemes using OLTP workloads (Section~\ref{sec:experiment}). 
\end{itemize}

\section{Evaluation}
\label{sec:eval}
In this section, we present the evaluation results to demonstrate the performance characteristics of PostgreSQL and MariaDB on different bandwidths. We aim to answer the following questions:
\begin{itemize}
    \item Is evaluation results from NVMeVirt paper reproducible in our environmental setup?
    (Section~\ref{sec:oltp})
    \item How do PostgreSQL and MariaDB act differently to bandwidth when running OLAP workload?
    (Section~\ref{sec:olap})
\end{itemize}

%% 엇 유진씨 이제 보니까 맥시멈 페이지 수 6페이지네요 !!!! 피규어 넣어도 안줄여도 될 수도 있겠네여 ㅎㅎㅎㅎ 
\subsection{Environmental Setup}
We used a Google Compute Engine instance running on Ubuntu 22.04 with kernel 6.1.14. The instance was equipped with one Intel Xeon processor operating at 2.20 GHz and has 24 cores and 128 GiB of memory in a NUMA configuration. For our evaluation, we dedicated 12 cores and 112 GiB of memory to NVMeVirt, and 12 cores and 16 GiB of memory were dedicated to the database engine.

To set up the environment as close as the environment used in the NVMeVirt paper, we established an NVMeVirt instance configured as an NVM SSD and set the I/O latency to a minimum. We configured the database instance with recommended settings obtained from optimization tools \cite{pgtune, mysqltuner}.

\begin{figure}[t]
    \centering
    \subfloat[TPS vs. target bandwidth]{\includegraphics[width=0.225\textwidth]
    {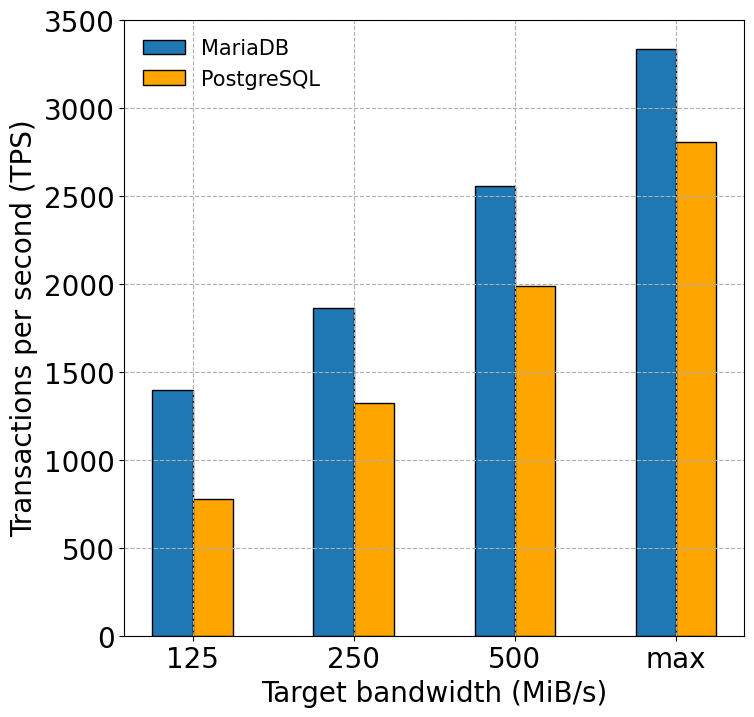}\label{fig:oltp_tps}}\hskip1.5ex
    \subfloat[I/O request]{\includegraphics[width=0.235\textwidth]{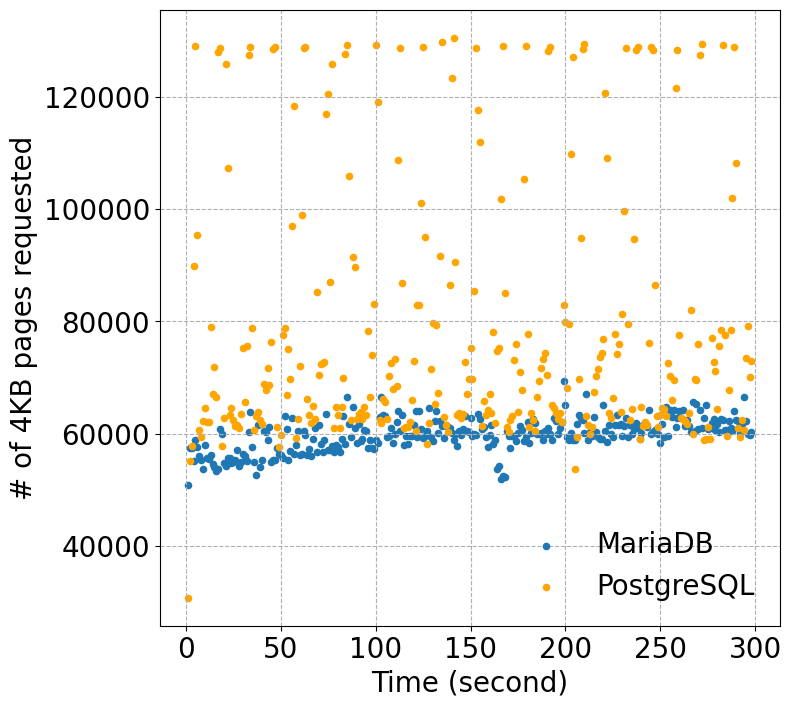}\label{fig:oltp_io}}
    \caption{Performance comparison on MariaDB and PostgreSQL on various bandwidth configurations with OLTP workload.}
\end{figure}

\begin{figure}
    \centering
    \subfloat[MariaDB]{\includegraphics[width=0.23\textwidth]{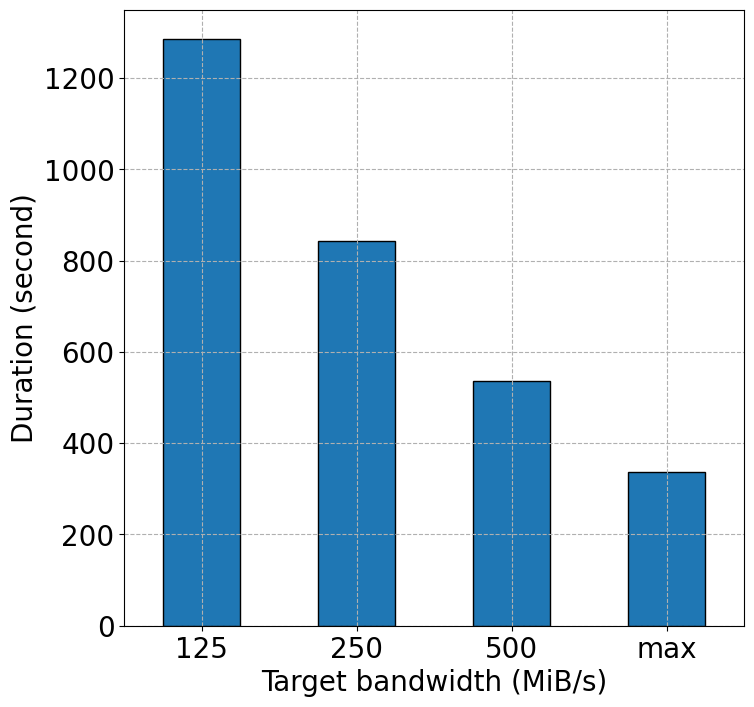}\label{fig:olap_maria}}\hskip1.5ex
    \subfloat[PostgreSQL]{\includegraphics[width=0.23\textwidth]{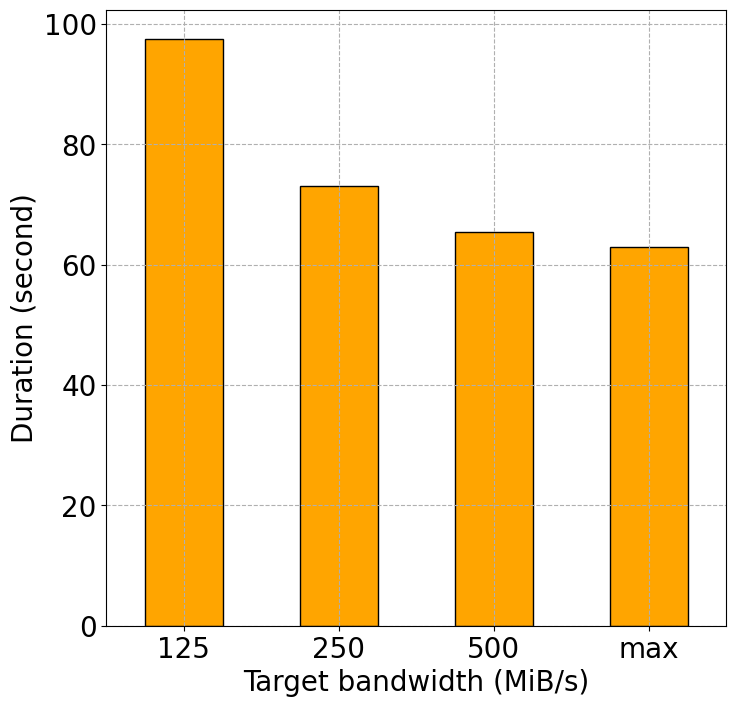}\label{fig:olap_psql}}
    \caption{Performance comparison on MariaDB and
PostgreSQL on various bandwidth configurations with OLAP workload.}
    \label{fig:olap_q18}
\end{figure}

\begin{figure}[t]
    \centering
    \includegraphics[width=0.6\linewidth]{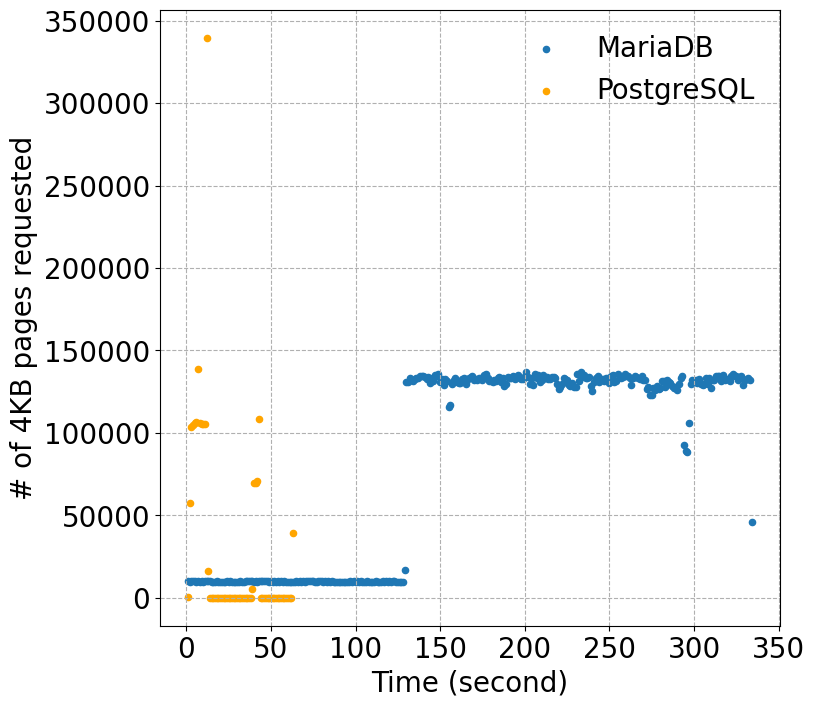}
    \caption{Number of 4KB pages requested by the MariaDB and PostgreSQL during OLAP query processing.}
    \label{fig:olap_io}
\end{figure}

\subsection{Online Transaction Processing Workload (OLTP)}
\label{sec:oltp}
We first compare the database engines' performance with the OLTP workload. We populated the database instance with sysbench to have ten tables of 25,000,000 bytes in size, a total size of approximately 60 GiB. Then we run the OLTP workload with sysbench with 12 threads for 30 minutes. We only report the first 5 minutes of trends since the performance became stable afterward.

Because the maximum bandwidth is bounded by the aggregated performance of the data processing units, the maximum target bandwidth in our environment is bounded by around 660 MiB/s. It's much smaller than that of NVMeVirt paper. Also, we set 12 CPU affinity to the database engine, whereas 36 cores were dedicated in the NVMeVirt paper. Due to these resource limitations, we could not produce the exact same result.

However, we still observed the three identical observations. Figure~\ref{fig:oltp_tps}  
compares processing performance measured in transactions per second (TPS) on various bandwidth limits. The first identical observation is that MariaDB outperforms PostgreSQL when the target bandwidth is low. In our evaluation, MariaDB outperforms PostgreSQL for every target bandwidth. Also, the performance gap becomes smaller as the target bandwidth increases. MariaDB outperforms PostgreSQL by 1.80x at a 125 MiB/s bandwidth limit and by 1.19x at maximum target bandwidth. This verifies that PostgreSQL is more sensitive to I/O bandwidth than MariaDB. Figure~\ref{fig:oltp_io} shows the number of 4KB pages the database engine requests to the device for each second. The device is configured to have a target bandwidth of a maximum, which is 660 MiB/s. This shows that PostgreSQL requires a higher number of pages, resulting in a higher number of disk I/O operations, even when the TPS is lower than that of MariaDB.

\subsection{Online Analytical Processing Workload (OLAP)}
\label{sec:olap}
We also compare the database engines' performance with the OLAP workload. We populated the database instance with TPC-H \cite{tpch} to have a 5 GiB dataset. Then we run the OLAP workload using 22 complex queries. We only report Query 18, which performs \verb|JOIN| on multiple original tables.

The result is very different from the OLTP workload. Figure~\ref{fig:olap_maria} and Figure~\ref{fig:olap_psql} compare the duration spent processing the query. PostgreSQL outperforms MariaDB significantly for every target bandwidth. The performance gap becomes smaller as the target bandwidth increases. PostgreSQL outperforms MariaDB by 13.18x at a 125 MiB/s bandwidth limit and by 5.30x at the maximum target bandwidth. Figure~\ref{fig:olap_io} shows the number of 4KB pages requested over time. MariaDB requires a higher number of disk I/O compared to PostgreSQL.

\subsection{Discussion}
OLTP and OLAP workload evaluation show opposite results. With OLTP workload, PostgreSQL performs worse when the storage is slow but it significantly outperforms MariaDB for OLAP workload regardless of I/O bandwidth. This hints that the differences in OLTP and OLAP workload processing scheme make PostgreSQL more sensitive to I/O bandwidth than MariaDB.

OLTP workload is write-focused with simple operations, while the OLAP workload is read-focused with complex operations \cite{workload}. Also, the OLTP workload involves a large number of concurrent transactions. Database management systems (DBMSs) maintain the illusion of isolation and, at the same time, interleave the operations of concurrent transactions for better performance. To decide the proper interleaving of operations, DBMSs employ {\itshape concurrency control}. These are the main differences between OLTP and OLAP workload processing.

\begin{table*}[t]
    \centering
    \begin{tabular}{c|c|c|c|c}
        \toprule 
        & Protocol & Version Storage & Garbage Collection & Index Management \\
        \hline
        PostgreSQL & MV2PL/SSI & Append-only (O2N) & Tuple-level (VAC)   & Physical Pointers \\
        MariaDB-InnoDB & MV2PL & Delta & Tuple-level (VAC) & Logical Pointers \\
        \bottomrule
    \end{tabular}
    \caption{MVCC Implementations \cite{MVCC}}
    \label{tab:mvcc}
\end{table*}

\section{Characteristics of Database Engine}
\label{sec:internal}
Based on the evaluation results, we have identified a potential explanation for the different performance characteristics: {\itshape Concurrency Control}. 
% CC 중에 I/O 많이 발생했다는 실험 결과가 없어서 빼야할 것 같습니다ㅠㅠ
% The experiments revealed significant disparities in the observed differences in I/O generation during concurrency control between the two DBMSs.
Concurrency control refers to the mechanisms employed by DBMSs to manage multiple transactions executing simultaneously, ensuring their isolation and preserving the consistency of the database \cite{CC}.
% These findings highlighted the fundamental distinction between OLTP and OLAP and the critical role of concurrency control in their respective workloads.
% OLTP workload에서의 CC의 역할만 강조하는게 더 좋을 것 같아요!
These findings highlighted the fundamental distinction between OLTP and OLAP and the critical role of concurrency control in OLTP workload.
% As a result, we redirected our focus toward analyzing the factors responsible for the variations in I/O generation during concurrency control in both PostgreSQL and MariaDB.
% 단어 I/O overhead로 통일했습니다!
As a result, we redirected our focus toward analyzing the factors responsible for the I/O overhead during concurrency control in PostgreSQL and MariaDB.

\subsection{Multi-Version Concurrency Control}
Both PostgreSQL and MariaDB employ a concurrency control method called multi-version concurrency control (MVCC). MVCC is widely adopted in modern relational DBMSs and encompasses concurrency control protocols, version storage, garbage collection, and index management \cite{MVCC}. 
% However, there are notable differences in the design decisions of Mariadb and PostgreSQL regarding each of these elements. In this study, we delve into the analysis of Version Storage, Garbage Collection, and Index Management, as these factors contribute to the disparity in I/O generation between the two DBMSs.
However, there are notable differences in the design decisions of MariaDB and PostgreSQL regarding each of these elements.

\subsection{Version Storage}
Under the MVCC system, a new physical version of the tuple is created when a transaction updates a tuple. The storage scheme employed by the DBMS determines how these versions are stored and the information contained in each version. The DBMS utilizes the pointer field of tuples to establish a version chain. 
% , which is implemented as a latch-free linked list.
This version chain enables the DBMS to locate the specific version of a tuple that is visible to a transaction.
% In the subsequent discussion, 
We provide a detailed explanation of these storage schemes, focusing on the trade-offs involved in \verb|UPDATE| operations.
% , as versioning is primarily handled within that context.
% The DBMS can insert new tuples into a table without requiring updates to other versions, and it deletes tuples by setting a flag in the begin-ts field of the current version. In the following sections, we explore the implications of these storage schemes on garbage collection and index pointer maintenance within the DBMS.

\begin{figure}[h]
    \centering
    \subfloat[Append-Only Storage]{\includegraphics[width=0.2\textwidth]{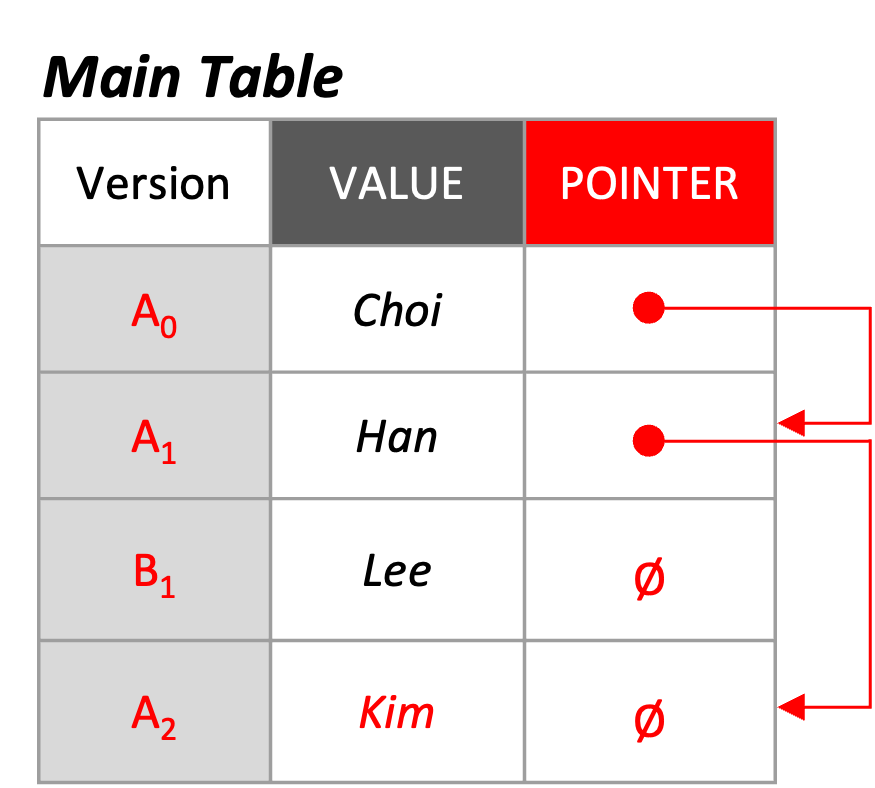}\label{fig:psql-ver}}\hskip1.5ex
    \centering
    \subfloat[Delta Storage]{\includegraphics[width=0.4\textwidth]{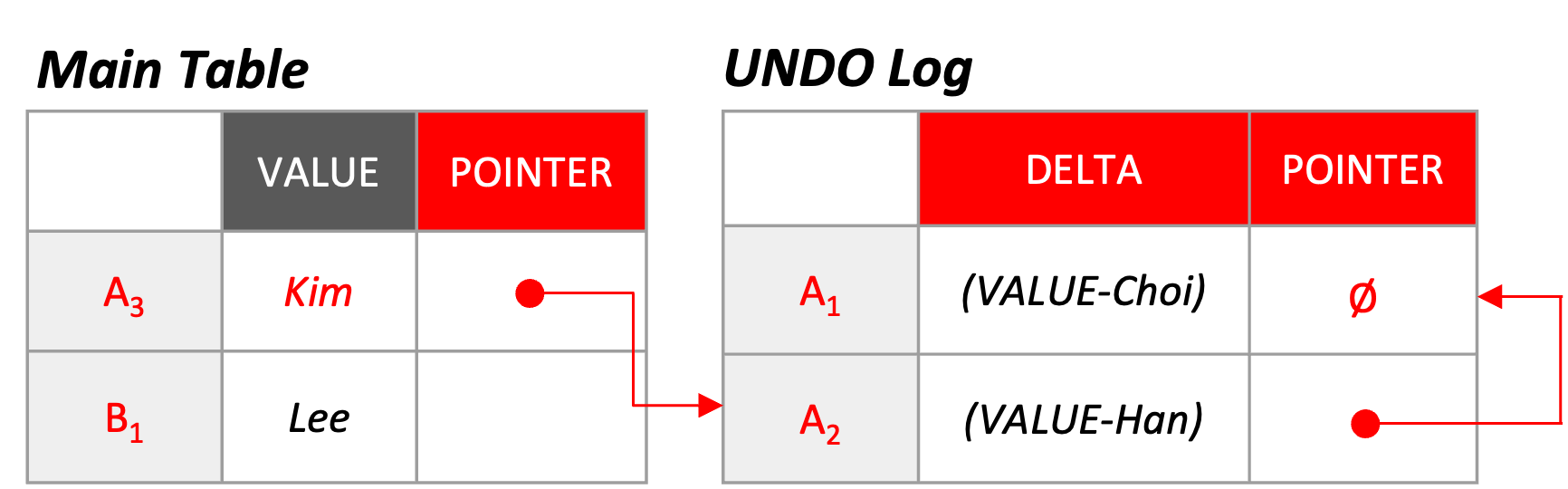}\label{fig:maria-ver}}
    \caption{PostgreSQL uses {\itshape append-only storage scheme} and MariaDB uses {\itshape delta storage scheme}.}
    \label{fig:olap_q18}
\end{figure}

\paragraph{Append-Only Storage}
PostgreSQL adopts the append-only approach as its version storage strategy \cite{internal}. In this strategy, all row versions of a table are stored in the same storage space. To modify an existing tuple, the DBMS follows a process where it allocates an empty slot from the table for the new version of the row. The content of the current version is then copied to the newly allocated slot, and the modifications are applied to this new version. Consequently, there are two physical versions representing the same logical row.

To maintain the lineage of these versions for future reference, MVCC DBMSs create a {\itshape version chain} using a singly linked-list structure. The version chain is unidirectional to minimize storage and maintenance overhead. The DBMS must determine the order in which the versions are organized: newest-to-oldest (N2O) or oldest-to-newest (O2N). While most DBMSs, such as Oracle and MySQL, implement N2O, 
% PostgreSQL stands out by employing the O2N approach.
PostgreSQL employs the O2N approach. In the N2O order, each tuple version points to its previous version, and the head of the version chain always points to the latest version. On the other hand, in the O2N order, each tuple version points to its newer version, and the head represents the oldest tuple version. The O2N approach eliminates the need for the DBMS to update indexes to point to a newer version of the tuple every time it is modified. However, it may take longer for the DBMS to locate the latest version during query processing, potentially requiring the traversal of a lengthy version chain. 

When a tuple is updated, the DBMS replicates all of its columns into the new version, regardless of whether the update affects a single column or all of them. This approach leads to significant data duplication and increased storage requirements. As a result, PostgreSQL requires more memory and disk space to store a database compared to other DBMSs.
% This results in slower queries and higher costs, particularly in cloud environments.

% 이 내용까진 필요 없을 거 같아서 일단 주석 처리했습니다! 넣어도 괜찮을 거 같긴해요
% There was an effort to modernize PostgreSQL's version storage implementation. EnterpriseDB initiated the zheap project \cite{zheap} in 2013, aiming to replace the append-only storage engine with delta versions. However, as of our knowledge cutoff in 2023, the project has not seen significant updates, and its progress appears to have stalled.

\paragraph{Delta Storage}
MariaDB-InnoDB implements a more efficient method called delta storage for version storage. Instead of duplicating the entire tuple for a new version, it stores a compact delta representing the changes between the new and current versions, similar to a \verb|git diff|. Consequently, when a query updates only a single column in a tuple with multiple columns, the DBMS only stores a delta record containing the specific change. The DBMS manages the main versions of tuples in the primary table and a series of delta versions in a separate storage called the UNDO log in MariaDB.

In MariaDB, the current version of a tuple resides in the primary table. When updating an existing tuple, the DBMS acquires a contiguous space from the delta storage to create a new delta version. This delta version solely includes the original values of the modified attributes, rather than duplicating the entire tuple. Subsequently, the DBMS directly performs an in-place update to the master version in the primary table. This approach proves advantageous for \verb|UPDATE| operations that modify only a subset of a tuple's attributes.

% 이건 in-memory 내용인 것 같아요
% since it minimizes memory allocations. 

% However, it incurs additional overhead for read-intensive workloads. To execute a read operation that accesses multiple attributes of a single tuple, the DBMS must traverse the version chain to retrieve the data for each accessed attribute. 이 내용을 넣는 게 좋을지? 안 넣는 것이 좋을듯요

\subsection{Garbage Collection}
If a MVCC-based DBMS doesn't reclaim unnecessary versions, the system will eventually experience storage space issues. This leads to longer query execution times as the DBMS must navigate lengthy version chains. Therefore, the performance of MVCC DBMSs greatly relies on the effectiveness of its garbage collection (GC) mechanism in safely reclaiming space during transactions. During the GC process, the DBMS performs three important steps:
\begin{enumerate}
    \item Detect expired versions of tuples.
    \item Unlink these versions from their associated chains and indexes.
    \item Reclaim the storage space occupied by these expired versions.
\end{enumerate}
These steps are essential for maintaining the optimal functioning of the DBMS and ensuring transactional safety.

Both PostgreSQL and MariaDB employ tuple-level GC called {\itshape vacuum} as part of their MVCC systems. This approach is crucial for managing storage space effectively and improving query performance. The vacuum involves assessing the visibility of individual tuples within the database. While both databases utilize vacuum, their mechanisms differ significantly due to variations in their version storage methods.

\begin{figure}[h]
    \centering
    \includegraphics[width=\linewidth]{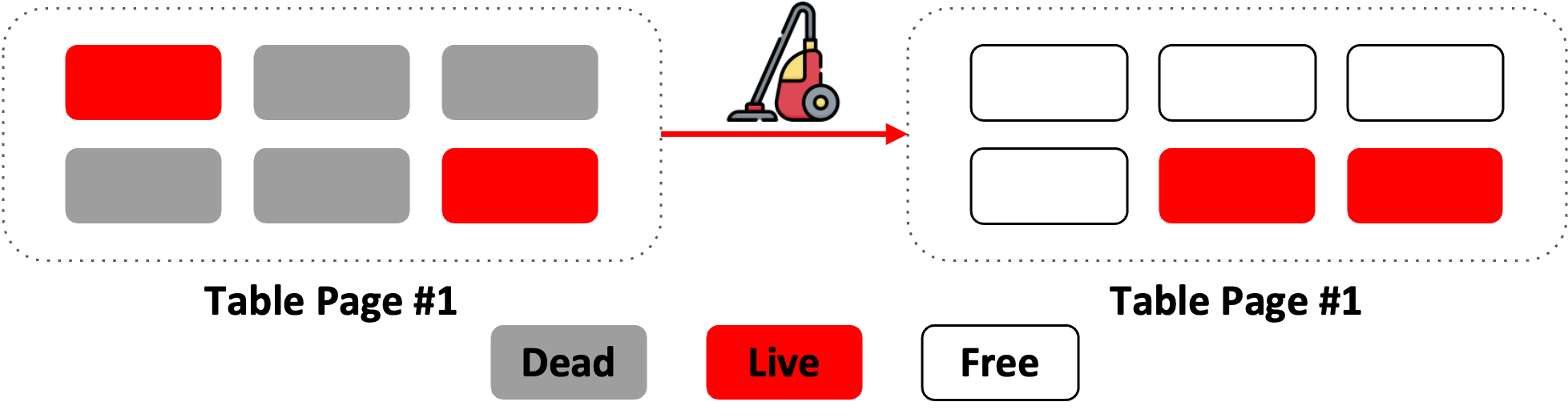}
    \caption{PostgreSQL vacuum process.}
    \label{fig:vacuum}
\end{figure}

\paragraph{Vacuum Management}
Figure~\ref{fig:vacuum} illustrates how the vacuum process in PostgreSQL works. PostgreSQL creates copies of rows during updates. The next aspect to consider is handling {\itshape dead tuples}, older versions that need to be removed. In the original PostgreSQL version from the 1980s, dead tuples were not removed to support "time-travel" queries, allowing examination of past versions. However, this resulted in tables not shrinking when tuples were deleted and long version chains for frequently updated tuples, impacting query performance. To address this, PostgreSQL adds index entries to quickly access the correct version, but it increases index size, affecting performance.

% 이것도 in-memory 내용
% and memory usage. These issues are interconnected.

In PostgreSQL, dead tuples occupy more space compared to delta versions. PostgreSQL uses the vacuum procedure to remove dead tuples. The procedure scans modified table pages since the last execution and identifies expired versions. An expired version is not visible to any active transaction and future transactions use the latest live version. Removing expired versions reclaims space for reuse, ensuring safety. PostgreSQL's {\itshape autovacuum} eventually removes dead tuples, but write-heavy workloads can cause accumulation faster than vacuuming, leading to continuous database growth.

% In-memory 내용 아닌지? 아리까리
% Dead tuples are loaded into memory during query execution, impacting query performance, IOPS, and memory usage during table scans. Inaccurate optimizer statistics caused by dead tuples can result in poor query plans.

% PostgreSQL automatically executes the vacuum procedure (autovacuum) based on system configuration settings. It can be further configured at the table level for fine-tuning. Users can also manually trigger the vacuum process using the VACUUM SQL command to optimize database performance.

MariaDB also uses a vacuum as its garbage collection mechanism. However, there is a difference in the areas where garbage collection occurs. Due to storing delta records in the undo log, MariaDB employs a {\itshape purge thread} that continuously scans the delta area in the background. When a transaction is committed, the purge thread removes the tuples stored in the undo log.

\subsection{Index Management}
MVCC-based DBMSs maintain a separation between versioning information and indexes. Index entries consist of key/value pairs, where the key represents the indexed attributes and the value is a pointer to the tuple. By following this pointer, the DBMS can access the tuple's version chain and find the version visible to a transaction. While false positive matches can occur, false negatives never happen with indexes.

Primary key indexes always point to the latest tuple version. In a delta scheme, the index points to the master version, reducing update frequency. In an append-only scheme, the index requires updates when new versions are created, such as when the primary key is modified.

\begin{figure}[h]
    \centering
    \subfloat[Physical Pointers]{\includegraphics[width=0.20\textwidth]{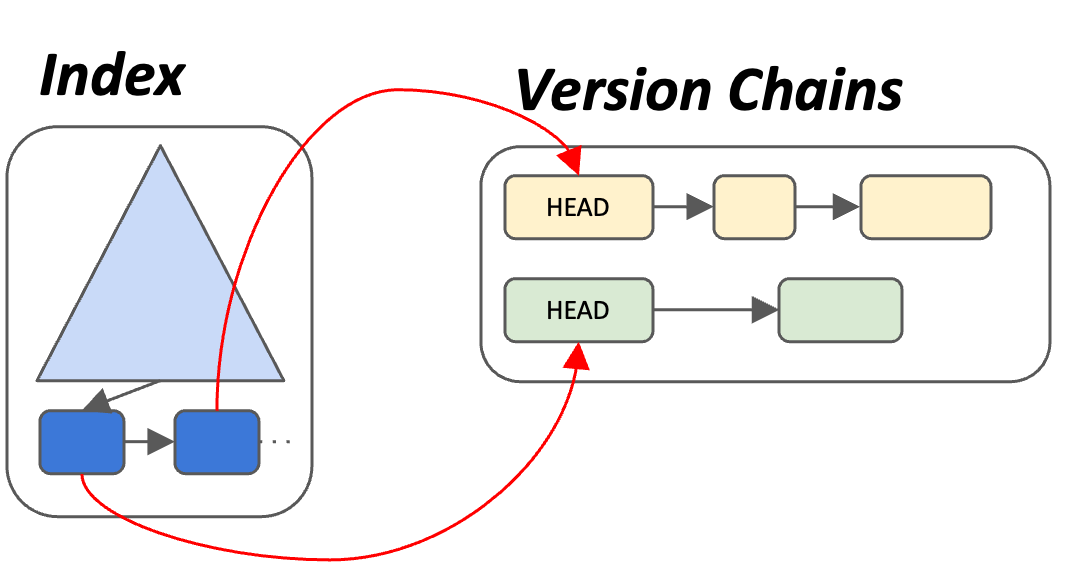}\label{fig:psql-idx}}\hskip2.5ex
    \centering
    \subfloat[Logical Pointers]{\includegraphics[width=0.24\textwidth]{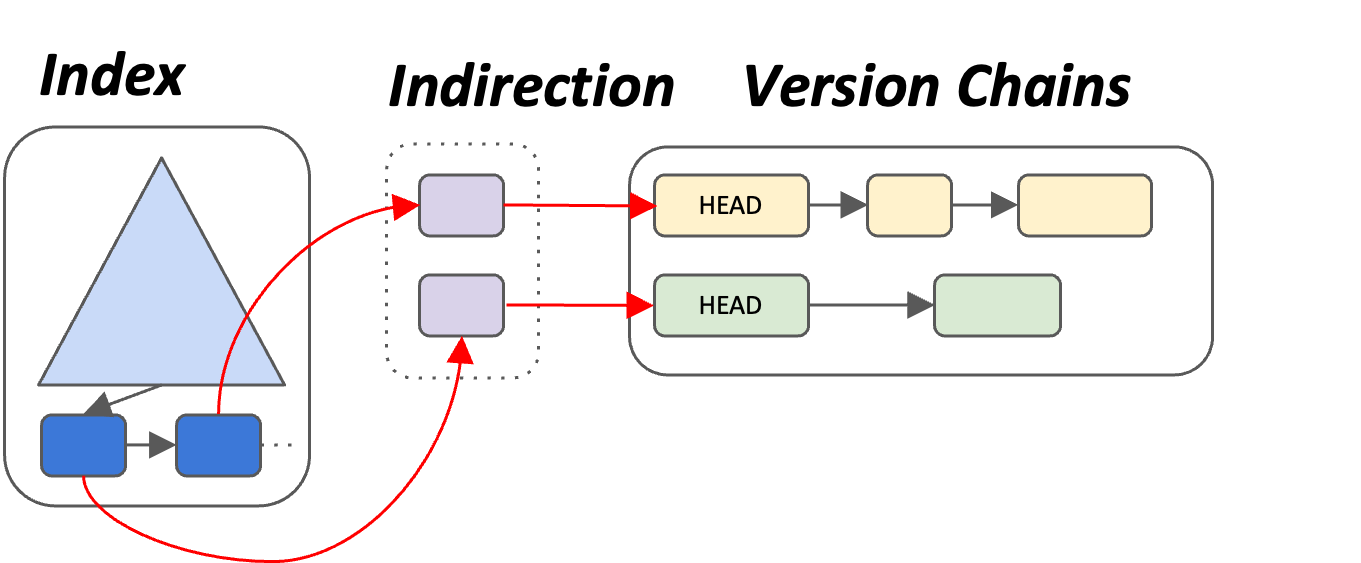}\label{fig:maria-idx}}
    \caption{PostgreSQL uses physical pointers and MariaDB uses logical pointers for secondary index management.}
    \label{fig:idx}
\end{figure}

Managing secondary indexes is more complex as both keys and pointers can change. Two approaches are used: physical pointers provide the exact tuple version's location as the index value, while logical pointers use indirection to map to the physical location. Figure~\ref{fig:idx} illustrates how PostgreSQL and MariaDB use pointers for secondary index management.

\paragraph{Physical Pointers}
PostgreSQL utilizes the physical pointers method, storing the physical addresses of versions in index entries \cite{internal}. This approach is specifically designed for append-only storage, allowing direct referencing of versions within the same table through indexes. When a tuple in a table is updated, the DBMS inserts the newly created version into all secondary indexes. This enables the DBMS to search for a tuple using a secondary index without comparing the secondary key against all indexed versions.

Each row's header in PostgreSQL contains a tuple ID field that points to the next version or its own tuple ID if it is the latest version. Therefore, when a query requests the latest version of a row, the DBMS traverses the index, starting from the oldest version and following the pointer until it reaches the desired version. However, this complete traversal of the version chain can be inefficient, especially when most queries only require the latest version. Consequently, update queries can become slower due to increased workload. The DBMS incurs additional I/O operations to traverse each index and insert new entries, leading to lock/latch contention in both the index and internal data structures like the buffer pool's page table. Notably, PostgreSQL performs this maintenance work for all indexes in a table, even if some queries may never utilize them. These extra reads and writes can pose challenges.

% , particularly in DBMSs that charge users based on IOPS, such as Amazon Aurora. Additionally, managing old versions and performing cleanup presents its own set of challenges.

To optimize disk I/O and minimize the need for multiple index entries and storing related versions across multiple pages, PostgreSQL employs a technique called {\itshape heap-only tuple} (HOT) updates \cite{hot}. When an update does not modify any columns referenced by the table's indexes and there is available space on the same data page as the old version, the DBMS creates a new copy of the tuple within the same disk page. This allows the index to still point to the old version, and queries retrieve the latest version by traversing the version chain. During regular operation, PostgreSQL further improves this process by removing old versions to prune the version chain.

\paragraph{Logical Pointers}
MariaDB adopts the logical pointers approach in index management \cite{mysql}. The key distinction lies in the architectural difference between the two. While PostgreSQL directly maps index records to on-disk locations, MariaDB employs a secondary structure. In MariaDB, secondary index records store a pointer to the primary key value instead of the on-disk row location, as done in PostgreSQL with the \verb|ctid|. MariaDB implements an indirection layer that maps a tuple's identifier to the head of its version chain. This design choice eliminates the need to update all indexes of a table to point to a new physical location whenever a tuple is modified, even if the indexed attributes remain unchanged. Only the mapping entry needs to be updated. However, since the index does not directly point to the exact version, the DBMS traverses the version chain starting from the HEAD to locate the visible version.

\begin{figure}
    \centering
    \subfloat[Read]{\includegraphics[width=0.23\textwidth]{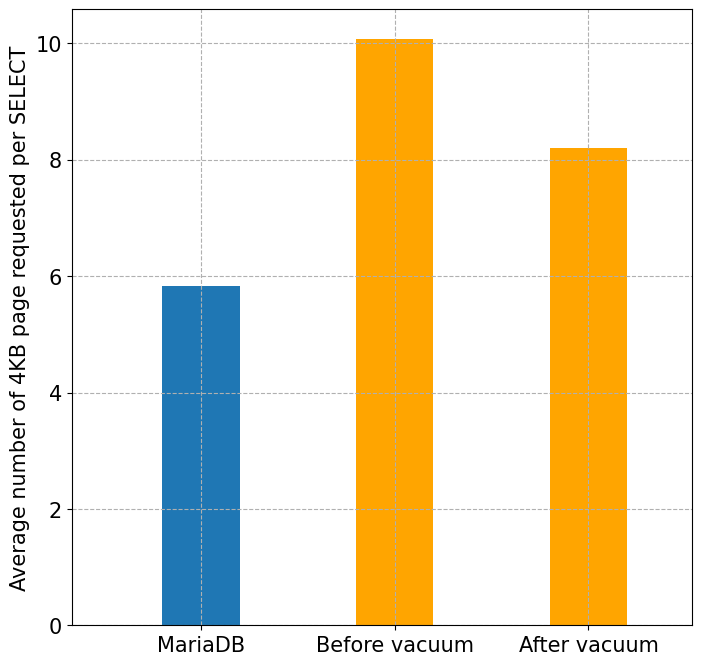}\label{fig:read}}\hskip1.5ex
    \subfloat[Write]{\includegraphics[width=0.23\textwidth]{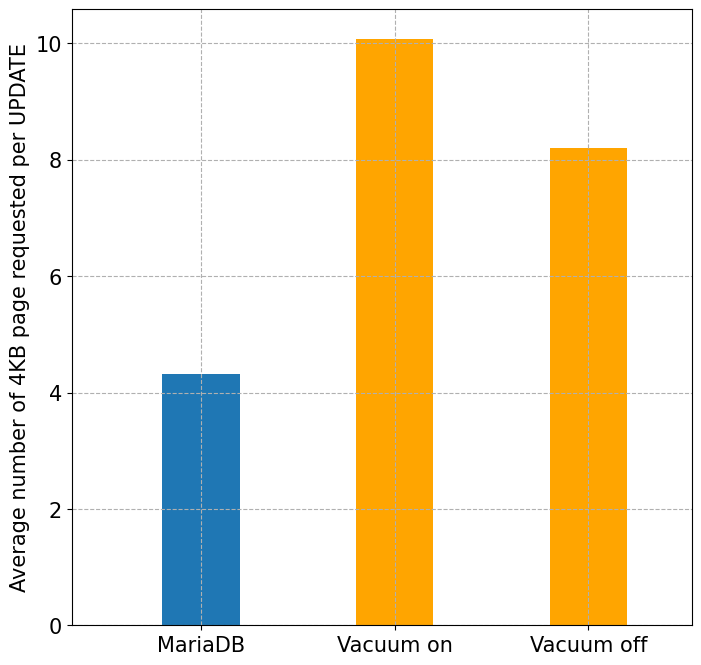}\label{fig:write}}
    \caption{Comparing I/O overhead for different MVCC approaches.}
    \label{fig:olap_q18}
\end{figure}

\section{Experimental Analysis}
\label{sec:experiment}
In this section, we present the experiment analysis to demonstrate the I/O overhead of PostgreSQL's MVCC scheme. For the experiment, we slightly modified the sysbench code and added a workload that only does \verb|UPDATE| (a {\itshape write} operation). Also, we used the workload that only does \verb|SELECT| (a {\itshape read} operation).

\subsection{Version Chain Scan Overhead}
To demonstrate the I/O overhead of scanning the version chain, we conducted an evaluation using the following procedures:

\begin{enumerate}
    \item Set PostgreSQL's autovacuum configuration to off so that all the multi versions of updated tuples remain in the table space.
    \item Run the workload that only does \verb|UPDATE|.
    \item Run the workload that only does \verb|SELECT|.
    \item Manually vacuum all the table spaces using \verb|VACUUM| command in PostgreSQL so that all the logical tuples can have only one physical tuple.
    \item Run the workload that only does \verb|SELECT|.
\end{enumerate}
MariaDB (more precisely, InnoDB) has no configuration that turns off the auto garbage collection. So we only run Step 2 and 3.

Figure~\ref{fig:read} visualizes the average number of 4KB pages requested per \verb|SELECT| query. Scanning a version chain requires approximately two more 4KB pages per read. Also, MariaDB requires approximately six 4KB pages per read which outperforms PostgreSQL even when the read happens after the vacuum. This shows that append-only storage results in higher I/O overhead for reads.

\subsection{Garbage Collection Overhead}
To demonstrate the I/O overhead of garbage collection, we conducted an evaluation. For PostgreSQL, we run the workload that only does \verb|UPDATE| with autovacuum configuration on and off. For MariaDB, we also run the workload that only does \verb|UPDATE|.

Figure~\ref{fig:write} visualizes the average number of 4KB pages requested per \verb|UPDATE| query. When autovacuum is on, PostgreSQL needs approximately two more 4KB pages per write. Also, MariaDB requires approximately four 4KB pages per write which outperforms PostgreSQL even when the autovacuum in PostgreSQL is off. This shows that, since the background vacuum process has to scan the table space that is interleaved with different versions of different tuples, it requires more disk I/O. Also, even when the autovacuum is off, the version chain needs to be scanned for writes since PostgreSQL uses the O2N scheme and results in higher number of I/O than the MariaDB.

\section{Future Work}
In our study, we could not observe the performance of the database engines with a bandwidth larger than 660 MiB/s. Therefore, conducting further analysis using higher resource capacities is important. This will allow us to better understand why PostgreSQL is a more promising database engine for future storage and MariaDB cannot fully utilizes the high I/O bandwidth. Also, it would be valuable to analyze the characteristics of OLAP workloads to investigate why PostgreSQL outperforms MariaDB in complex queries involving operations such as \verb|JOIN| and \verb|AGGREGATION|. This will provide insights into the specific features and optimizations the PostgreSQL uses.

\section{Conclusion}
We presented an analysis of database engine internals that explains why PostgreSQL utilizes the storage device more eagerly than MariaDB. We provide an explanation for different MVCC implementations that PostgreSQL and MariaDB uses. We also conducted the experimental analysis to evaluate the I/O overhead for MVCC design decisions.

%% 7 section으로 references하고 싶은데 저 밑에 레퍼런스는 어떻게 없애는 것인지,...
%% 이게 ACM paper format의 자체 형식인거라 수정이 안돼요ㅠㅠ
\bibliographystyle{ACM-Reference-Format}
\bibliography{references}


\begin{thebibliography}{13}

%%% the Web address.

\ifx \showCODEN    \undefined \def \showCODEN     #1{\unskip}     \fi
\ifx \showDOI      \undefined \def \showDOI       #1{#1}\fi
\ifx \showISBNx    \undefined \def \showISBNx     #1{\unskip}     \fi
\ifx \showISBNxiii \undefined \def \showISBNxiii  #1{\unskip}     \fi
\ifx \showISSN     \undefined \def \showISSN      #1{\unskip}     \fi
\ifx \showLCCN     \undefined \def \showLCCN      #1{\unskip}     \fi
\ifx \shownote     \undefined \def \shownote      #1{#1}          \fi
\ifx \showarticletitle \undefined \def \showarticletitle #1{#1}   \fi
\ifx \showURL      \undefined \def \showURL       {\relax}        \fi
% invisible to TeX
\providecommand\bibfield[2]{#2}
\providecommand\bibinfo[2]{#2}
\providecommand\natexlab[1]{#1}
\providecommand\showeprint[2][]{arXiv:#2}

\bibitem[Barghouti and Kaiser(1991)]%
        {CC}
\bibfield{author}{\bibinfo{person}{Naser~Saleh Barghouti} {and}
  \bibinfo{person}{Gail~Elaine Kaiser}.} \bibinfo{year}{1991}\natexlab{}.
\newblock \showarticletitle{Concurrency Control in Advanced Database
  Applications}. In \bibinfo{booktitle}{\emph{ACM Computing Surveys}},
  Vol.~\bibinfo{volume}{23}.
\newblock
Issue 3.


\bibitem[Foundation(2023)]%
        {mariadb}
\bibfield{author}{\bibinfo{person}{MariaDB Foundation}.}
  \bibinfo{year}{2023}\natexlab{}.
\newblock \bibinfo{booktitle}{\emph{MariaDB}}.
\newblock
\urldef\tempurl%
\url{https://mariadb.org}
\showURL{%
\tempurl}


\bibitem[Group(2023a)]%
        {hot}
\bibfield{author}{\bibinfo{person}{PostgreSQL Global~Development Group}.}
  \bibinfo{year}{2023}\natexlab{a}.
\newblock \bibinfo{booktitle}{\emph{Heap-only Tuple}}.
\newblock
\urldef\tempurl%
\url{https://www.postgresql.org/docs/current/storage-hot.html}
\showURL{%
\tempurl}


\bibitem[Group(2023b)]%
        {psql}
\bibfield{author}{\bibinfo{person}{PostgreSQL Global~Development Group}.}
  \bibinfo{year}{2023}\natexlab{b}.
\newblock \bibinfo{booktitle}{\emph{PostgreSQL}}.
\newblock
\urldef\tempurl%
\url{https://www.postgresql.org}
\showURL{%
\tempurl}


\bibitem[Hayden(2023)]%
        {mysqltuner}
\bibfield{author}{\bibinfo{person}{Major Hayden}.}
  \bibinfo{year}{2023}\natexlab{}.
\newblock \bibinfo{booktitle}{\emph{MySQLTuner}}.
\newblock
\urldef\tempurl%
\url{https://github.com/major/MySQLTuner-perl}
\showURL{%
\tempurl}


\bibitem[Kim et~al\mbox{.}(2023)]%
        {NVMeVirt:FAST23}
\bibfield{author}{\bibinfo{person}{Sang-Hoon Kim}, \bibinfo{person}{Jaehoon
  Shim}, \bibinfo{person}{Euidong Lee}, \bibinfo{person}{Seongyeop Jeong},
  \bibinfo{person}{Ilkueon Kang}, {and} \bibinfo{person}{Jin-Soo Kim}.}
  \bibinfo{year}{2023}\natexlab{}.
\newblock \showarticletitle{{NVMeVirt}: A Versatile Software-defined Virtual
  {NVMe} Device}. In \bibinfo{booktitle}{\emph{Proceedings of the 21st USENIX
  Conference on File and Storage Technologies (USENIX FAST)}}.
  \bibinfo{address}{Santa Clara, CA}.
\newblock


\bibitem[Kopytov(2023)]%
        {sysbench}
\bibfield{author}{\bibinfo{person}{Alexey Kopytov}.}
  \bibinfo{year}{2023}\natexlab{}.
\newblock \bibinfo{booktitle}{\emph{sysbench}}.
\newblock
\urldef\tempurl%
\url{https://github.com/akopytov/sysbench}
\showURL{%
\tempurl}


\bibitem[Oracle(2023)]%
        {mysql}
\bibfield{author}{\bibinfo{person}{Oracle}.} \bibinfo{year}{2023}\natexlab{}.
\newblock \bibinfo{booktitle}{\emph{MySQL}}.
\newblock
\urldef\tempurl%
\url{http://www.mysql.com}
\showURL{%
\tempurl}


\bibitem[Solutions(2023)]%
        {tpch}
\bibfield{author}{\bibinfo{person}{Hotea Solutions}.}
  \bibinfo{year}{2023}\natexlab{}.
\newblock \bibinfo{booktitle}{\emph{TPC-H}}.
\newblock
\urldef\tempurl%
\url{https://www.tpc.org/tpch}
\showURL{%
\tempurl}


\bibitem[Stonebraker and Cattell(2011)]%
        {workload}
\bibfield{author}{\bibinfo{person}{Michael Stonebraker} {and}
  \bibinfo{person}{Rick Cattell}.} \bibinfo{year}{2011}\natexlab{}.
\newblock \showarticletitle{10 Rules for Scalable Performance in ‘Simple
  Operation’ Datastores}.
\newblock \bibinfo{journal}{\emph{Commun. ACM}} \bibinfo{volume}{54},
  \bibinfo{number}{6}, \bibinfo{pages}{72--80}.
\newblock


\bibitem[Suzuki(2023)]%
        {internal}
\bibfield{author}{\bibinfo{person}{Hironobu Suzuki}.}
  \bibinfo{year}{2023}\natexlab{}.
\newblock \bibinfo{booktitle}{\emph{The Internals Of PostgreSQL}}.
\newblock
\urldef\tempurl%
\url{https://www.interdb.jp/pg/}
\showURL{%
\tempurl}


\bibitem[Vasyliev(2023)]%
        {pgtune}
\bibfield{author}{\bibinfo{person}{Oleksii Vasyliev}.}
  \bibinfo{year}{2023}\natexlab{}.
\newblock \bibinfo{booktitle}{\emph{PGTune}}.
\newblock
\urldef\tempurl%
\url{https://pgtune.leopard.in.ua}
\showURL{%
\tempurl}


\bibitem[Wu et~al\mbox{.}(2017)]%
        {MVCC}
\bibfield{author}{\bibinfo{person}{Yingjun Wu}, \bibinfo{person}{Joy Arulraj},
  \bibinfo{person}{Jiexi Lin}, \bibinfo{person}{Ran Xian}, {and}
  \bibinfo{person}{Andrew Pavlo}.} \bibinfo{year}{2017}\natexlab{}.
\newblock \showarticletitle{An Empirical Evaluation of In-memory Multi-version
  Concurrency Control}.
\newblock \bibinfo{journal}{\emph{Proc. of the VLDB Endowment}}
  \bibinfo{volume}{10}, \bibinfo{number}{7}, \bibinfo{pages}{781--792}.
\newblock


\end{thebibliography}

\end{document}